\documentclass[conference]{IEEEtran}
\IEEEoverridecommandlockouts
\def\BibTeX{{\rm B\kern-.5em{\sc i\kern-.024em b}\kern-.09em
		T\kern-.166em\lower.7ex\hbox{E}\kern-.12emX}}

\usepackage{etoolbox}
\usepackage{soul}
\usepackage{color}
\usepackage{filecontents,lipsum}
\usepackage[noadjust]{cite}
\usepackage{amsthm,amssymb,amsmath,bm}
\usepackage{amsfonts}
\usepackage{epsfig}
\usepackage{epstopdf}
\usepackage{amssymb}
\usepackage{amsmath}
\usepackage{cite}
\usepackage[Algorithm]{algorithm}
\usepackage{color,soul}
\usepackage{multirow}
\usepackage{rotating}
\usepackage{graphicx}
\usepackage{tabularx}
\usepackage{array}
\usepackage{color,soul}
\usepackage{graphicx,dblfloatfix}
\usepackage{blindtext}
\usepackage[dvipsnames]{xcolor}
\usepackage{amsthm,amssymb,amsmath,bm}
\usepackage{graphicx}
\usepackage{fancyhdr}
\usepackage[font={small}]{caption}
\usepackage{subfig}
\usepackage{tabularx}
\usepackage{cite}
\usepackage[utf8]{inputenc}
\usepackage[T1]{fontenc}
\usepackage{authblk}

\begin{document}
\title{Cloud-based Queuing Model for Tactile Internet in Next Generation of RAN}
\author[1]{Narges Gholipoor}
\author[2]{Saeedeh Parsaeefard}
\author[3]{Mohammad Reza Javan}
\author[1]{Nader Mokari}
\author[1]{Hamid Saeedi}
\author[4]{\\Hossein Pishro-Nik}
\affil[1]{Department of Electrical and Computer Engineering, Tarbiat Modares University, Iran}
\affil[2]{Iran Telecommunication Research Center, Iran}
\affil[3]{Department of Electrical and Robotic Engineering, Shahrood University of Technology, Iran}
\affil[4]{Department of Electrical and Computer Engineering, University of Massachusetts, Amherst, USA}
\renewcommand\Authands{, and }

\makeatletter
\patchcmd{\@maketitle}

\makeatother


%


\maketitle
\begin{abstract}
Ultra-low latency is the most important requirement of the Tactile Internet (TI), which is one of the proposed services for the next-generation wireless network (NGWN), e.g., fifth-generation (5G) network.  In this paper, a new queuing model for the TI is proposed for the cloud radio access network (C-RAN) architecture of the NGWN by applying power domain non-orthogonal multiple access (PD-NOMA) technology.
In this model, we consider both the radio remote head (RRH) and baseband processing unit (BBU) queuing delays for each end-to-end (E2E) connection between a pair of tactile users.
  In our setup, to minimize the transmit power of users subject to guaranteeing an acceptable delay of users, and fronthaul and access constraints, we formulate a resource allocation (RA) problem. Furthermore, we dynamically set the fronthaul and access links to minimize the total transmit power.
  Given that the proposed RA problem is highly non-convex, in order to solve it, we utilize diverse transformation techniques such as successive convex approximation (SCA) and difference of two convex functions (DC).
     Numerical results show that by dynamic adjustment of the access and fronthaul delays, transmit power reduces in comparison with the fixed approach per each connection.
       Also, energy efficiency of orthogonal frequency division multiple access (OFDMA) and PD-NOMA are compared for our setup.
\end{abstract}
\begin{IEEEkeywords}
	\vspace{-0.5em}
C-RAN, Tactile Internet, NOMA.
\end{IEEEkeywords}
\IEEEpeerreviewmaketitle
\vspace{-0.5em}
	\section{Introduction}
	\vspace{-0.5em}
In the NGWNs, e.g., the fifth-generation (5G) wireless networks,  new services such as The Tactile Internet (TI) is introduced.
 One of the most critical requirements of {TI} services is ultra-low end-to-end (E2E)  delay, e.g., E2E delay of about $1$~ms \cite{aijaz2017realizing,fettweis2014tactile,she2016ensuring,simsek20165g}.  This delay requirement cannot be ensured via existing wireless networks such as fourth-generation wireless networks (4G)  \cite{simsek20165g}.
 However, 5G platform  via its own  virtualized, slice-based, and cloud-based architecture can realize the implementation the {TI} services  \cite{aijaz2017realizing,fettweis2014tactile}.
  
By utilizing cloud radio access network (C-RAN) in 5G, spectral efficiency (SE), and energy efficiency (EE) along with cost can be efficiently optimized, where baseband processing is performed by the baseband units (BBUs) which are connected to remote radio heads (RRHs) via the fronthaul links \cite{park2016joint,7064897}.
   Moreover, in a dense environment, C-RAN decreases energy consumption, cost, and increases throughput and SE \cite{zhang2016fronthauling}.  Hence, C-RAN is a suitable architecture for the implementation of the {TI} services in dense areas.

One of the promising approaches for the NGWN is the power domain non-orthogonal multiple access (PD-NOMA)\footnote{We use PD-NOMA and NOMA interchangeably in the followings.} which can significantly improve the SE and reduce the transmission delay \cite{6692652}. In NOMA, the available spectrum is shared among different users of one RRH. A NOMA transmitter requires implementing a method for superposition of different users signals. Consequently, to compensate interference between users, a complex method for decoding the signals is required at a NOMA receiver. Successive interference cancellation (SIC) is one of these methods which is implemented at the receiver to decode the desirable signal  \cite{7676258,8115155}. 
		

In the NGWN,  various services with different requirements are realized with a high quality of service (QoS) with the aid of the virtualization techniques. Hence, the concept of the slicing is defined for each type of services in which each slice is a bundle of  resources for users with a specific set of QoS requirements 
	 \cite{7926923,8004168}. 
Slicing provides flexibility to utilize resources which improve SE and EE. However,  in order that the activities of users of one slice do not have detrimental effects on the QoS of the users of other slices, the isolation between slices should be maintained.
	  There exists various literature for translation to isolation concept to the proper notation for the networks’ procedures such as dynamic and static methods \cite{7462252,7926923,8004168}. In this paper, we consider the minimum required data rate of each slice for preserving the isolation between slices  \cite{7462252}.

In \cite{she2016ensuring}, cross-layer RA problem for the {TI} is investigated for single-cell network where the packet error probability, maximum allowable queuing delay violation probability, and packet dropping probability are jointly optimized to minimize the total transmit power subject to maximum allowable queuing delays.
 In \cite{aijaz2016towards}, queuing delay, queuing delay violation, packet error, and packet drop resulting from channel fading are considered for investigating the E2E delay of RAN.
A cross-layer RA problem for the {TI} in the  Long-Term Evolution-Advanced (LTE-A) is studied in which the queuing delay and queuing delay violation in one base station (BS) are optimized in \cite{aijaz2016towards}. Orthogonal frequency division multiple access (OFDMA)  and single carrier frequency division multiple access (SC-FDMA) are considered for downlink (DL) and uplink (UL), respectively. 
In \cite{she2016uplink}, the effect of frequency diversity and spatial diversity on the transmission reliability in UL is studied in the {TI} service, where the number of subcarriers and the bandwidth of each subcarrier are optimized for minimizing the total bandwidth to guarantee the transmission reliability.
In \cite{8253477}, a multi-cell network based on frequency division multiple access (FDMA) with a fixed delay for backhaul is studied in the {TI} service. 

 In all the above works, one queue at the BS is considered for each user. Hence, by increasing the number of users in each cell, a lot of queues are required at the BS. since the {TI} is expected to realize in the 5G framework, it should be investigated in the C-RAN architecture. 
Furthermore, previous works consider orthogonal multiple access techniques instead of NOMA. Moreover, all of the above works ignore the fronthaul delay. However, on account of the importance of delay in the {TI}, it is essential to consider fronthaul queuing delay, otherwise, the result of RA may not meet the requirements of {TI} services.
To tackle these issues, we investigate a C-RAN architecture with a set of paired tactile users. The contributions of this paper are as follows, many of which have been considered for the first time in the {TI}:
\vspace{-0.25em}
\begin{itemize}
	\item
	We propose a C-RAN scenario in ultra-dense environment in 5G platform. 
	Moreover, for this setup, we propose a practical queuing model 
	that can be implemented in realistic networks for {TI}  services. 
  We consider a PD-NOMA scheme for our system model while all earlier works are based on OFDMA and FDMA schemes. Moreover, we consider slicing for the {TI} service in our work. 
	\item \textcolor{black}{In contrast to \cite{she2016ensuring,she2016uplink,aijaz2016towards} where the fronthaul delay is ignored, we take this delay into consideration. Moreover,
	we consider dynamic adjustment of the access and fronthaul delays based on channel state information (CSI) for each pair of users instead of fixed maximum delay values  per each transmission part of our setup and show that it can significantly reduce the required total transmit power.}	
\end{itemize}
The rest of this paper is as follows. In Section II, the system model is described. In Section III, we formulate the optimization problem. In Section IV, we illustrate the iterative algorithm for solving problem formulation. Numerical results and simulation are presented in Section V. Finally, Section VI concludes the paper. 
\vspace{-0.5em}
\section{System Model}
\textcolor{black}{We consider a C-RAN network where all RRHs are connected to the BBU via fronthaul links.
 In this region, there exist several pairs of tactile users  where each user aims to send its information to  its paired tactile user.} 
 \textcolor{black}{Assume each RRH has only one queue for UL transmission and all the data of tactile users is stored in this queue. RRHs send all of the data to the BBU to process. Then, the BBU sends the data to the corresponding RRH. In DL,  we assume that each RRH has a queue for each user for sending data to the paired users.}
 \begin{figure}[t] \label{picS}
 	\centering 
 	\includegraphics[width=0.5\textwidth]{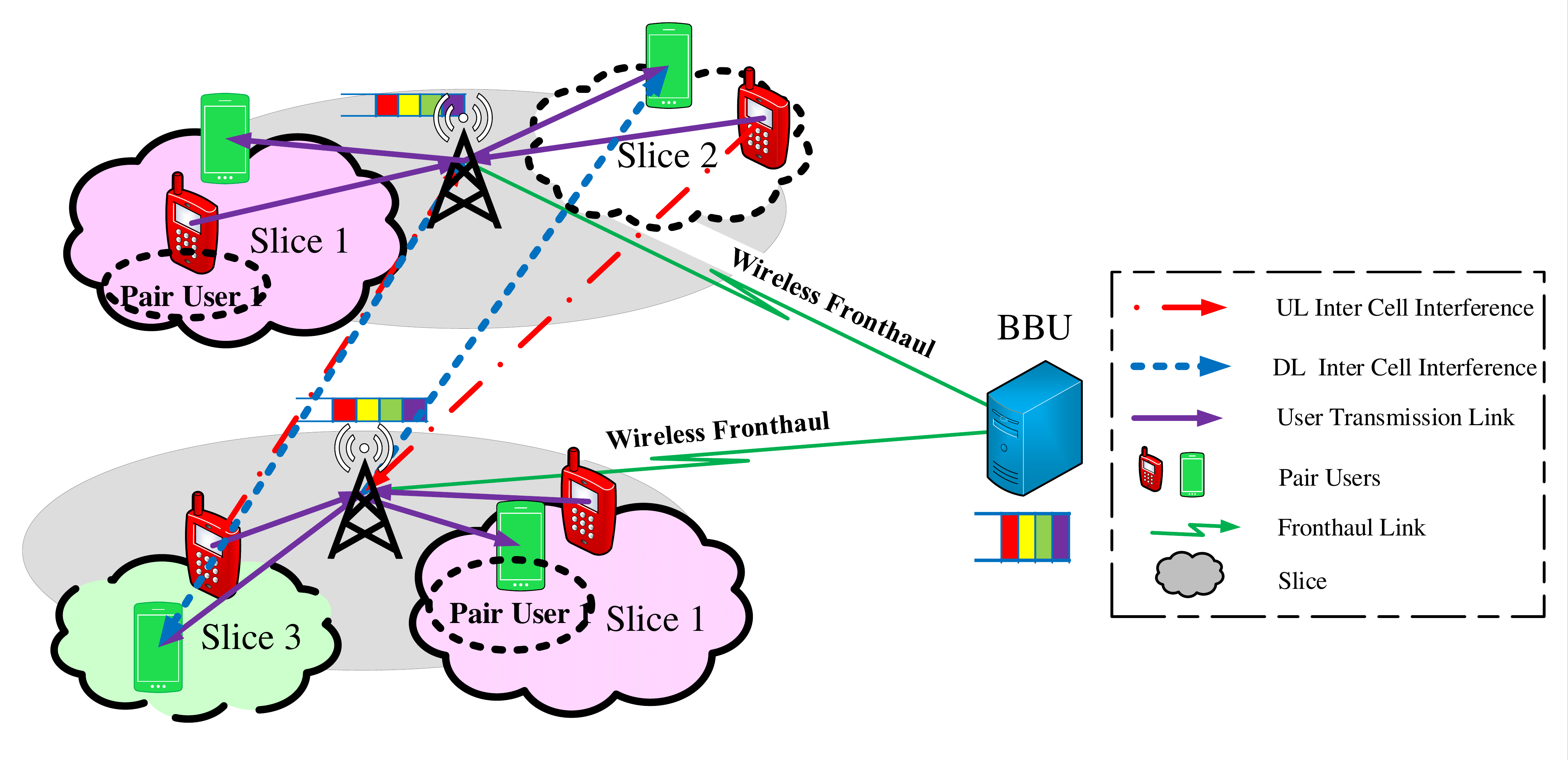}
 	\vspace{-2em}
  		\caption{The illustration of the considered network in which three slices with two RRHs are considered. Here as an example, a pair of tactile users is shown by the dotted circles.}
  		 	\vspace{-1em}
  	\label{pic}
 \end{figure}
As shown in Fig. \ref{pic}, in the proposed system model, we consider  $\boldsymbol{\mathcal{J}}=\{1,\dots,J\}$ RRHs, $\boldsymbol{\mathcal{S}}=\{1,\dots,S\}$ slices, and  $\boldsymbol{\mathcal{I}}=\{0,...,I\}$ pairs of tactile users. Slice $s$ contains $\boldsymbol{\mathcal{I}_s}=\{0,...,I_s\}$ tactile users and the total number of tactile users in our system model is equal to $\boldsymbol{\mathcal{I}}=\bigcup_{s\in \mathcal{S}}{\boldsymbol{\mathcal{I}_s}}$ pairs of users.
It worth noting that the access link and fronthaul link denote to the RRH-user link and RRH-BBU link, respectively.
  In the C-RAN architecture to reduce the cabling cost, wireless fronthaul is utilized instead of fiber fronthaul  \cite{park2016joint}. In this paper, we consider a C-RAN platform for an ultra-dense network with wireless fronthaul and that there exist two sets of subcarriers $\boldsymbol{\mathcal{K}_1}=\{1,\dots,K_1\}$  and $\boldsymbol{\mathcal{K}_2}=\{1,\dots,K_2\}$ for access and fronthaul links, respectively. 
  \textcolor{black}{Moreover, we define $\mathcal{Q}=\{\text{UL},\text{DL}\}$ for simplicity. We consider a two-phase transmission; in the first phase, all tactile users  send their data to the corresponding RRH and simultaneously all RRHs send their buffered data to the BBU via fronthaul links.	In the second phase, all RRHs send data to the corresponding tactile users , and simultaneously, BBU sends  the buffered data to all RRHs via fronthaul links.}
\textcolor{black}{Therefore, the proposed system model is based on the frequency division duplex  (FDD) mode in which each RRH can transmit and receive simultaneously in different frequencies.} In order to isolate slices, a minimum required data rate for each slice $s$ must be reserved \cite{7462252}. 
 \vspace{-0.5em}
\subsection{Access Links Parameters}
We introduce a binary variable $\tau_{i,k_1}^{s,j,q}$ which is set to 1 if subcarrier $k_1$ is assigned to user $i$ in slice $s$ at RRH $j$.
Since we deploy PD-NOMA in this setup, each subcarrier can be allocated to $L_1$ users \cite{7676258}. Therefore, we have the following constraint:
\vspace{-0.5em}
\begin{equation} \label{eqo2}
{\text{C1: }}\sum_{s \in \mathcal{S}}\sum_{i \in \mathcal{I}_s}\tau_{i,k_1}^{s,j,q} \le L_1, \forall j\in\mathcal{J}, k_1\in\mathcal{K}_1, q\in\mathcal{Q}.\nonumber
\end{equation}
Here, for all $i\in\mathcal{I}$, $j\in\mathcal{J}$, $k_1\in\mathcal{K}_1$, $s\in\mathcal{S}$ and $q\in\mathcal{Q}$, the achievable rate for user $i$ on subcarrier $k_1$ at RRH $j$ can be calculated as
$
r_{i,k_1}^{s,j,q}= \log_2(1+\gamma^{s,j,q}_{i,k_1}),
$
where  
$\gamma^{s,j,q}_{i,k_1}=\frac{p_{i,k_1}^{s,j,q}h_{i,k_1}^{s,j,q}}{\sigma_{i,k_1}^{s,j,q}+I_{i,k_1}^{s,j,q}+\tilde I_{i,k_1}^{s,j,q}}$, in which $p_{i,k_1}^{s,j,q}$, $h_{i,k_1}^{s,j,q}$, and $\sigma_{i,k_1}^{s,j,q}$ represent the transmit power, channel power gain from  RRH $j$ to user $i$  on subcarrier $k_1$ in slice $s$, and noise power, respectively. Also, $ I_{i,k_1}^{s,j,q}$ is intra-cell interference which is equal to $
I_{i,k_1}^{s,j,q}=\sum_{v\in\mathcal{S}}\sum_{\substack{ u\in\mathcal{I}_v,u\neq i\\h_{u,k_1}^{s,j,q}> h_{i,k_1}^{v,j,q}}} \tau_{u,k_1}^{v,j,q}p_{u,k_1}^{v,j,q}h_{i,k_1}^{s,j,q},
$
and $\tilde I_{i,k_1}^{s,j,q}$  is inter-cell interference which is equal to 
$\tilde I_{i,k_1}^{s,j,q}=\sum_{f\in\mathcal{J},f\neq j}\sum_{v\in\mathcal{S}}\sum_{u\in\mathcal{I}_{v}} \tau_{u,k_1}^{v,f,q}p_{u,k_1}^{v,f,q}h_{i,k_1}^{v,f,q}
$. Therefore, the total achievable rate in the access links at RRH $j$ is as follows:
\begin{equation} \label{eqo7}
R_{\text{RRH}_j}^{q}=\sum_{s\in\mathcal{S}}\sum_{u\in \mathcal{I}_s}\sum_{k_1\in\mathcal{K}_1}\tau_{u,k_1}^{s,j,q} r_{u,k_1}^{s,j,q} \forall  j\in\mathcal{J}, q\in\mathcal{Q}.
\end{equation}

Due to the power limitation of each RRH in DL transmission, we have the following constraint:
\begin{equation} \label{eqo71}
{\text{C2: }} \sum_{i\in \mathcal{I}} \sum_{s\in \mathcal{S}} \sum_{k_1\in \mathcal{K}_1}\tau_{i,k_1}^{s,j,\text{DL}} p_{i,k_1}^{s,j,\text{DL}}\le P_{\text{RRH}_j}^{\text{DL}}, \forall  j\in\mathcal{J}.\nonumber
\end{equation}	
Moreover, due to the power limitation of each user, we have
\begin{equation} \label{eqo72}
{\text{C3: }}\sum_{j\in \mathcal{J}} \sum_{s\in \mathcal{S}} \sum_{k_1\in \mathcal{K}_1}\tau_{i,k_1}^{s,j,\text{UL}} p_{i,k_1}^{s,j,\text{UL}}\le P_{\text{USER}_i}^{\text{UL}}, \forall  i\in\mathcal{I}.\nonumber
\end{equation}

Due to using NOMA in access, we have SIC constraint as follows:
\vspace{-0.5em}
\begin{equation}
\begin{array}{l}
{\text{C4: }}
\frac{{\tau_{i,{k_1}}^{{s,j},q}p_{m',{k_1}}^{{s,j},q}h_{i,{k_1}}^{{s,j},q}}}{{\tilde I_{i,{k_1}}^{{s,j},q} + I_{i,{k_1}}^{{s,j},q} + \sigma _{i,{k_1}}^{{s,j},q}}} \ge \frac{{\tau_{i,{k_1}}^{{s,j},q}p_{m',{k_1}}^{{s,j},q}h_{m',{k_1}}^{{s,j},q}}}{{\tilde I_{m',{k_1}}^{{s,j},q} + I_{m',{k_1}}^{{s,j},q} + \sigma _{m',{k_1}}^{{s,j},q}}},\\ \forall j, {s}, q, i,m' \in \mathcal{I}, m' \neq i, h^{s,j, q}_{i,k_1} > h^{s,j, q}_{m',k_1}.
\end{array}\nonumber
\end{equation}
\vspace{-1.5em}
\subsection{Fronthaul Links Parameters}
We introduce a binary variable $x_{k_2}^{j,q}$ which is set to 1 if subcarrier $k_2$ is assigned to RRH $j$, otherwise, set to 0.
Assuming that PD-NOMA is also deployed for the fronthaul links, again each subcarrier can be allocated to at most  $L_2$ RRHs, and hence, we have the following constraint:
\begin{equation} \label{eqo9}
{\text{C5: }}\sum_{j \in \mathcal{J}}x_{k_2}^{j,q} \le L_2,\forall k_2\in\mathcal{K}_2,q\in\mathcal{Q} .\nonumber
\end{equation}
The achievable rate for each RRH on subcarrier $k_2$ is calculated as $
r_{k_2}^{j,q}=\log_2(1+\gamma_{k_2}^{j,q}), \forall j\in\mathcal{J}, k_2\in\mathcal{K}_2,q\in\mathcal{Q} ,
$, 
where  $\gamma_{k_2}^{j,q}$ is defined as 
$
\gamma_{k_2}^{j,q}=\frac{p_{k_2}^{j,q}h_{k_2}^{j,q}}{\sigma_{k_2}^{j,q}+I_{k_2}^{j,q}}, \forall j\in\mathcal{J}, k_2\in\mathcal{K}_2,q\in\mathcal{Q},
$
where $ I_{k_2}^{j,q}$ is the interference among RRHs and is represented as $
I_{k_2}^{j,q}=\sum_{\substack{f\in\mathcal{J},\\f\neq j,h_{k_2}^{f,q}> h_{k_2}^{j,q}}} x_{k_2}^{f,q}p_{k_2}^{f,q}h_{k_2}^{j,q}, \forall j\in\mathcal{J}, k_2\in\mathcal{K}_2,q\in\mathcal{Q}$. Therefore, the total achievable rate in the BBU is obtained as follows:
\vspace{-0.5em}
\begin{equation} \label{eqo13}
R_{\text{BBU}}^{q}= \sum_{j\in \mathcal{J}}\sum_{k_2\in\mathcal{K}_2} x_{k_2}^{j,q}r_{k_2}^{j,q}, \forall q\in\mathcal{Q}.
\end{equation}
Due to the power limitation of each RRH in UL transmission, we have
\vspace{-0.5em}
\begin{equation} \label{eqo71}
{\text{C6: }} 
 \sum_{k_2\in \mathcal{K}_2} x_{k_2}^{j, \text{UL}} p_{k_2}^{j,\text{UL}}\le P_{\text{RRH}_j}^{ \text{UL}},\forall  j\in\mathcal{J}.\nonumber
\end{equation}	
Moreover, due to the power limitation of the BBU, we have
\begin{equation} \label{eqo72}
{\text{C7: }}  \sum_{j\in \mathcal{J}} \sum_{k_2\in \mathcal{K}_2} x_{k_2}^{j,\text{DL}} p_{k_2}^{j,\text{DL}}\le P_{\text{BBU}}^{\text{DL}}.\nonumber
\end{equation}
Due to using NOMA in fronthaul, we have SIC constraint as follows:
\vspace{-1em}
\begin{equation}
\begin{array}{l}
{\text{C8: }}
\frac{{x_{{k_2}}^{{j},q}p_{{k_2}}^{{j'},q}h_{{k_2}}^{{j},q}}}{{I_{{k_2}}^{{j},q} + \sigma _{{k_2}}^{{j},q}}} \ge \frac{{x_{{k_2}}^{{j},q}p_{{k_2}}^{{j'},q}h_{{k_2}}^{{j'},q}}}{{I_{{k_2}}^{{j'},q} + \sigma _{{k_2}}^{{j'},q}}},\\\forall {j},q \in {\cal Q}, \forall j,j' \in \mathcal{J}, j' \neq j, h^{j, \mathcal{Q}}_{k_2} > h^{j', q}_{k_2}.
\end{array}\nonumber
\end{equation}
\vspace{-2em}
\subsection{Queuing Delay Model}
The total delay of this architecture consists of three components: delay resulting from UL queues at RRH, BBU queue, and DL  queues at RRHs.
Due to delay constraint in the {TI} service, we have
\begin{equation}
{\text{C9: }}D_{\text{max}}^j+D_{\text{max}}^{i,j}+D_{\text{max}}^{BBU}\le D^{\text{Total}}_{{\rm{max}}}, \forall i \in \mathcal{I},j\in \mathcal{J}, s\in \mathcal{S},\nonumber
\end{equation}
where $ D_{\text{max}}^j$, $D_{\text{max}}^{i,j}$, $D_{\text{max}}^{BBU}$, and $D^{\text{Total}}_{{\rm{max}}}$ are delays of UL queues  at RRH, BBU queue, DL queues at RRH, and total delay, respectively.
\subsubsection{UL Queuing Delay}
The aggregation of receiving bits from several nodes can be modeled as a Poisson process \cite{li2007network,she2016ensuring}. The effective bandwidth for a Poisson arrival process in RRH $j$ is defined as $
E_B^{j}(\theta_j)=\lambda_j \frac{(e^{\theta_j}-1)}{\theta_j}, \forall j\in\mathcal{J}$, 
where $\theta_j$ is the statistical QoS exponent of the $j^{\text{th}}$ RRH \cite{li2007network,she2016ensuring}. A larger  $\theta_j$ indicates a more stringent QoS and a smaller  $\theta_j$ implies a looser QoS requirement. $\lambda_j$ is the number of bits arrived at RRH $j$ queue defined as 
$
{\lambda _j} = \sum\limits_{{s} \in {{\cal S}}} {\sum\limits_{u \in {{\cal I}_s}} {\sum\limits_{{k_1} \in {{\cal K}_1}} {r_{u,{k_1}}^{{s,j},\text{UL}}} } },~ \forall j\in\mathcal{J}$.
 The probability of queuing delay violation for RRH $j$ can be approximated as
 \vspace{-0.5em}
\begin{equation} \label{deqo1756}
\begin{split}
\epsilon_1^{j}= \Pr\{D_j>D^j_{\text{max}}\}=\eta_1 \exp(-\theta_j E_B^{j}(\theta_j) D^j_{\text{max}}),
\end{split}
\vspace{-0.5em}
\end{equation}
for all $j\in\mathcal{J} $ where $D_j$ is the $j^{\text{th}}$ RRH delay, $D^j_{\text{max}}$ is the maximum delay, and $\eta_1$ is the  non-empty buffer probability. Equation \eqref{deqo1756} can be simplified to
\vspace{-0.5em}
\begin{equation}\nonumber
\begin{array}{l}
	\exp ( - {\theta _j}E_B^j({\theta _j}){D_{{\rm{max}}}}) = \exp ( - {\theta _j}{\lambda _j}\frac{{({e^{{\theta _j}}} - 1)}}{{{\theta _j}}}{D^j_{{\rm{max}}}})=\\ \exp (- {\lambda _j}({e^{{\theta _j}}} - 1){D^j_{{\rm{max}}}}) \le {\delta _1}.
\end{array}
\end{equation}
\vspace{-0.5em}
Therefore, we have	
\begin{equation}
\begin{split}
{\text{C10: }}\sum\limits_{{s} \in {{\cal S}}} {\sum\limits_{u \in {{\cal I}_s}} {\sum\limits_{{k_1} \in {{\cal K}_1}} {r_{u,{k_1}}^{{s,j},\text{UL}}} } }  \ge \frac{{\ln ({1/\delta _1})}}{{( {e^{{\theta _j}}}-1){D^j_{{\rm{max}}}}}}, \forall j \in \mathcal{J}.\nonumber
\end{split}
\end{equation}	
\vspace{-1 em}	
\subsubsection{BBU Queuing Delay}We consider a queue for all RRHs at the BBU for processing data. Therefore, the formulas in the previous section can also be used for this section.
The effective bandwidth for each queue in BBU is $
E_B^{\text{BBU}}(\theta_{\text{BBU}})=\Lambda_{\text{BBU}} \frac{(e^{\theta_{\text{BBU}}}-1)}{\theta_{\text{BBU}}}, 
$
where $\theta_{\text{BBU}}$ is the statistical QoS exponent in the BBU and $\Lambda_{\text{BBU}}$ is the number of bits arrived at  the queue in the BBU which is defined as $
{\Lambda_{\text{BBU}}} ={\sum\limits_{{j} \in {{\cal J}}}}{\sum\limits_{{k_2} \in {{\cal K}_2}} {r_{{k_2}}^{{j},\text{UL}}} }.
$
The probability of queuing delay violation at the BBU can be approximated as
$\epsilon_{\text{BBU}}= \Pr\{D_{\text{BBU}}>D^{\text{BBU}}_{\text{max}}\}=\eta_2 \exp(-\theta_{\text{BBU}}^* E_B^{\text{BBU}}(\theta_{\text{BBU}}) D^{\text{BBU}}_{\text{max}}),
$
where $\eta_2$ is the non-empty buffer probability and this equation can be simplified to
\begin{equation}\nonumber
\begin{array}{l}
	\exp(-\theta_{\text{BBU}}^* E_B^{\text{BBU}}(\theta_{\text{BBU}}) D^{\text{BBU}}_{\text{max}})=\\ \exp(-\theta_{\text{BBU}}^* \Lambda_{\text{BBU}} \frac{(e^{\theta_{\text{BBU}}}-1)}{\theta_{\text{BBU}}} D^{\text{BBU}}_{\text{max}}) \le {\delta _2}.
\end{array}
\end{equation}
Therefore, we have
\vspace{-1em}	
\begin{equation}
\begin{split}
{\text{C11: }}{\sum\limits_{{j} \in {{\cal J}}}}{\sum\limits_{{k_2} \in {{\cal K}_2}} {r_{{k_2}}^{{j},\text{UL}}} } \ge \frac{{\ln ({1/\delta _2})}}{{(e^{\theta_{\text{BBU}}}-1){D^{\text{BBU}}_{{\rm{max}}}}}}.\nonumber
\end{split}
\end{equation}
\subsubsection{DL Queuing Delay}
The effective bandwidth for each user in  RRH $j$ is defined as
$
E_B^{i,j}(\theta_i^j)=\lambda_i^j \frac{(e^{\theta_i^j}-1)}{\theta_i^j},~\forall i\in\mathcal{I}, j\in\mathcal{J}, 
$
where $\theta_i^j$ is the statistical QoS exponent of the $i^{\text{th}}$ user in RRH $j$ and  $\lambda_i^j$ is the number of bits arrived at user $i$ queue in RRH $j$ which is defined as $
{\lambda_i^j} = \sum\limits_{{s} \in {{\cal S}}} \sum\limits_{{u} \in {{\cal I}_s}} {\sum\limits_{{k_1} \in {{\cal K}_1}} {r_{u,{k_1}}^{{s,j},\text{DL}}} }, \forall i\in\mathcal{I}, j\in\mathcal{J}$.
The probability of queuing delay violation for user $i$ can be approximated as
\begin{equation} \label{ddeqo17}
\begin{array}{l}
\epsilon_3^{i,j}= \Pr\{D_i^j>D^{i,j}_{\text{max}}\}=\\\eta_3 \exp(-\theta_i^j E_B^{i,j}(\theta_i^j) D_{\text{max}}), \forall i\in\mathcal{I}, j\in\mathcal{J}, 
\end{array}
\end{equation}
where $D_i^j$ is the $i^{\text{th}}$ user delay in RRH $j$ and $\eta_3$ is the  non-empty buffer probability. Equation \eqref{ddeqo17} can simplified to
\begin{equation}\nonumber
\begin{array}{l}
\exp ( - \theta _i^jE_B^{i,j}(\theta _i^j){D^{i,j}_{{\rm{max}}}}) = \exp ( - \theta _i^j\lambda _i^j\frac{{({e^{\theta _i^j}} - 1)}}{{\theta _i^j}}{D_{{\rm{max}}}}) =\\ \exp ( - \lambda {_i^j}({e^{\theta _i^j}} - 1){D^{i,j}_{{\rm{max}}}}) \le \delta_3.
\end{array}
\end{equation}
Therefore, we have
\vspace{-1em}	
\begin{equation}
\begin{split}
{\text{C12: }}\sum\limits_{{k_1} \in {\mathcal{K}_1}} {r_{u,{k_1}}^{s,j,\text{DL}}}  \ge \frac{{\ln (1/\delta_3 )}}{{({e^{\theta _i^j}} - 1){D^{i,j}_{{\rm{max}}}}}}, \forall i\in \mathcal{I}, \forall j \in \mathcal{J}.\nonumber
\end{split}
\end{equation}
In order to avoid bit dropping, the output rate of queues must be greater than the input rate of queues. Therefore, we have two following constraints:
\begin{equation}
{\text{C13: }}  {\sum\limits_{{s} \in {{\cal S}}}  \sum\limits_{u \in {\cal I}_s} {\sum\limits_{j \in {\cal J}} {\sum\limits_{{k_1} \in {{\cal K}_1}} {\tau_{u,k_1}^{s,j,\text{UL}} r_{u,{k_1}}^{{s,j},UL}}}}}\le{\sum\limits_{j \in {\cal J}} {\sum\limits_{{k_2} \in {{\cal K}_2}} {x_{k_2}^{j,\text{UL}}r_{{k_2}}^{{j},{\rm{UL}}}} } },\nonumber
\end{equation}
\begin{equation}
{\text{C14: }}{\sum\limits_{j \in {\cal J}} {\sum\limits_{{k_2} \in {{\cal K}_2}} {x_{k_2}^{j,\text{DL}}r_{{k_2}}^{{j},{\rm{DL}}}} }}\le {\sum\limits_{{s} \in {{\cal S}}}\sum\limits_{u \in {\cal I}_s}{\sum\limits_{j \in {\cal J}} {\sum\limits_{{k_1} \in {{\cal K}_1}} {\tau_{u,k_1}^{s,j,\text{DL}}r_{u,{k_1}}^{{s,j},{\rm{DL}}}} } } }.\nonumber
\end{equation}
\vspace{-0.5em}
\section{Optimization Problem Formulation} 
In this section, our aim is to allocate resources to minimize the overall power consumption in our setup by considering a bounded delay constraint to satisfy the E2E delay requirements. Based on the mentioned constraints C1-C14, the optimization problem can be written as
\vspace{-0.5em}
	\begin{align}\label{eqoa}
	&\min_{\boldsymbol{P},\boldsymbol{T},\boldsymbol{X},\boldsymbol{D}} \sum_{j\in \mathcal{J}}\sum_{k_2\in\mathcal{K}_2}\sum_{s\in\mathcal{S}}\sum_{u\in \mathcal{I}_s}\sum_{k_1\in\mathcal{K}_1}\sum_{q\in\mathcal{Q}}x_{k_2}^{j,q} p_{k_2}^{j,q} +\tau_{u,k_1}^{s,j,q} p_{u,k_1}^{s,j,q}\nonumber
	\\\text{s.t.}:&\text{(C1)-(C14)}, \\ &\text{C15: }\sum\limits_{{k_1} \in {{\cal K}_1}} {\sum\limits_{u \in {\cal I}_s} {\sum\limits_{j \in {\cal J}} {\tau_{u,k_1}^{s,j,q}r_{u,{k_1}}^{{s,j},q}} } }  \ge R^{s, q}_{{\rm{rsv}}},\forall s \in {\cal S},q \in {\cal Q}.~ \nonumber
		\vspace{-1em}
	\end{align}
The optimization variables in \eqref{eqoa} are subcarrier and power allocation for different users in access and fronthaul as well as in both UL and DL where $\boldsymbol{P}$, $\boldsymbol{T}$, $\boldsymbol{X}$, and $\boldsymbol{D}$ are the transmit power, the access subcarrier allocation, fronthaul subcarrier allocation, and delay vector for users, respectively. C15 is the rate constraint for isolation of slices. \textcolor{black}{According to the delay and SIC constraints in the optimization problem, one of the outputs of the optimization problem is the pair of NOMA users that satisfy SIC and delay constraints.
	In order to reduce complexity, we assume $L_1=L_2=2$. If these constraints are not met for two distinct users, each subcarrier is exclusively allocated to  at most one user.}
In problem \eqref{eqoa}, the rate is a non-convex function, which leads to the non-convexity of the problem. In addition, this problem contains both discrete and continuous variables, which makes the problem more challenging. Therefore, we resort to an alternate method to propose an efficient iterative algorithm \cite{ngo2014joint} with three sub-problems, namely, subcarrier allocation sub-problem, power allocation sub-problem, and delay adjustment sub-problem which will be explained in the followings. 
\vspace{-1.5em}
\section{An Efficient Iterative Algorithm}
As mentioned earlier, to solve \eqref{eqoa}, we deploy an iterative algorithm that divides the problem into three sub-problems and solve them alternately \cite{ngo2014joint}. This procedure is presented in Algorithm.\ref{ALG1}. Let $z$ be the iteration number and $\boldsymbol{P}^{(0)}$, $\boldsymbol{X}^{(0)}$, and $\boldsymbol{T}^{(0)}$ be the initial values. In each iteration, we solve each sub-problem with considering the optimization parameters of other sub-problems as fixed values derived in the previous steps. The iteration stops when the error in Step 5 is less than a predetermined threshold, i.e., $\epsilon_{\text{TH}}$, or the number of iterations exceeds a predetermined value i.e., $Z_{\text{TH}}$. The solution of the last iteration is then declared as the solution of \eqref{eqoa}.  To solve all of these subproblems, we use the difference of two convex (DC)  functions to transform the problem into a convex form. The transformed problem is a convex problem and can be solved with the CVX toolbox in Matlab \cite{grant2008cvx,boyd2004convex}. Due to page limitation, we provide more detail of solution method, the proof of convergence of algorithm and subproblems in the extended version of this paper \cite{gholipoor2019resource}.
\begin{algorithm}[t]
	\caption{Three-Step Iterative Algorithm}	
	
	{\textbf{Step 1: Initialization}}~\\ $\mathcal{J} = \{1,...,J\}$,$\mathcal{K}_1 = \{1,...,K_1\}$, $\mathcal{K}_2 = \{1,...,K_2\}$, $\mathcal{I}_s = \{1,...,I_j\}$, $\mathcal{S} = \{1,...,S\}$, $\epsilon_{\text{TH}}=10^{-4}$, $Z_{\text{TH}}=100$, $z=0$.\\ Calculate initial value $p^{(z)}=p^0$, $\tau^{(z)}=\tau^0$,  $x^{(z)}=x^0$.
	\\ {\textbf{Step 2: Subcarrier Allocation:}}
	Allocate subcarrier by minimizing the transmit power  and satisfying the problem constraints i.e, \eqref{optimizationtotalsub-carrierllocation11}.
	\\ {\textbf{Step 3: Power Allocation:}}
	Allocate power to each user according to problem (\ref{optimizationtotalpowerllocation1}) and subcarrier allocated in Step 2.
	\\ {\textbf{Step 4: Delay Adjustment:}}
	Adjust delay of each user  according to problem (\ref{deqoa}).
	\\ {\textbf{Step 5: Iteration}}
	$z=z+1$, Repeat Step 2, 3 and 4 until $||\boldsymbol{P}^{(z)}-\boldsymbol{P}^{(z-1)}||\le\epsilon_{\text{TH}}$ or  $Z_{\text{TH}}<z$.
	\label{ALG1}
\end{algorithm}
\subsection{Subcarrier Allocation Sub-Problem}
With assuming fixed value $\boldsymbol{P}$ and $\boldsymbol{D}$, the subcarrier allocation sub-problem is written as follows:
	\begin{align} \label{optimizationtotalsub-carrierllocation11} 
	&\min_{\boldsymbol{T},\boldsymbol{X}} \sum_{j\in \mathcal{J}}\sum_{k_2\in\mathcal{K}_2}\sum_{s\in\mathcal{S}}\sum_{u\in \mathcal{I}_s}\sum_{k_1\in\mathcal{K}_1}\sum_{q\in\mathcal{Q}}x_{k_2}^{j,q} p_{k_2}^{j,q} +\tau_{u,k_1}^{s,j,q} p_{u,k_1}^{s,j,q}\nonumber
	\\\text{s.t.}:&\text{(C1-C8), (C10-C15).} 
	\end{align}

While \eqref{optimizationtotalsub-carrierllocation11} has less computational complexity than \eqref{eqoa}, it suffers from  non-convexity due to the interference in the rate functions. In addition, this problem contains discrete variables. We apply time sharing method and relax discrete variables as $x_{k_2}^{j,q}\in[0,1],\forall k_2\in\mathcal{K}_2,\forall j \in \mathcal{J}, \forall q\in\mathcal{Q}$ and $\tau_{u,k_1}^{s,j,q}\in[0,1],\forall u\in\mathcal{I}_s, \forall k_1\in\mathcal{K}_1,\forall j \in \mathcal{J}, \forall q\in\mathcal{Q}$.
\vspace{-0.5em}
\subsection{Power Allocation Sub-Problem}
\vspace{-0.5em}
For the fixed value of $\boldsymbol{T}$, $\boldsymbol{X}$ and $\boldsymbol{D}$, the power allocation sub-problem is obtained as follows
\vspace{-0.5em}
	\begin{align}
	\label{optimizationtotalpowerllocation1}
	&\min_{\boldsymbol{P}} \sum_{j\in \mathcal{J}}\sum_{k_2\in\mathcal{K}_2}\sum_{s\in\mathcal{S}}\sum_{u\in \mathcal{I}_s}\sum_{k_1\in\mathcal{K}_1}\sum_{q\in\mathcal{Q}}x_{k_2}^{j,q} p_{k_2}^{j,q} +\tau_{u,k_1}^{s,j,q} p_{u,k_1}^{s,j,q}\nonumber
		\\\text{s.t.}:& \text{(C2)-(C4), (C6)-(C8), (C9)-(C15)}. 
	\end{align}
\vspace{-0.2em}
Similar to the subcarrier allocation sub-problem,  in problem \eqref{optimizationtotalpowerllocation1}, the rate is a non-convex function, which leads to the non-convexity of the problem. Therefore, it is necessary to approximate \eqref{optimizationtotalpowerllocation1} with a convex problem. To solve this problem, we use the DC approximation to transform the problem into a convex form. 
\vspace{-0.5em}
\subsection{Delay Adjustment Sub-Problem}
With assuming fixed values of $\boldsymbol{P}$, $\boldsymbol{T}$, and $\boldsymbol{X}$, the delay adjustment sub-problem is obtained as
\vspace{-0.5em}
	\begin{align}\label{deqoa}
	&\text{find}  \,\,\,\,\, {\boldsymbol{D}} 
	\\		\text{s.t.}:& \text{C9-C12}. \nonumber
	\end{align}
The delay adjustment sub-problem can be solved by the linear programming (LP) of any optimization toolbox.
\section{Simulation and Results}
In this section, the simulation results are presented to evaluate the performance of the proposed system model. To simulate dense urban area, we consider a BBU is at the center of the coverage area whose distance is $1$~Km from a set of RRHs. The coverage area is considered $10$ square Kilometers. Moreover, we consider a Rayleigh fading wireless channel in which the subcarrier gains are independent. Channel power gains for the access links are set as $h_{i,k_1}^{s,j,q}=\Omega_{i,k_1}^{s,j,q}{d_{i}^{j,q}}^{-\alpha}$ where $d_{i}^{j,q}$ is the distance between user $i$ and RRH $j$, $\Omega_{i,k_1}^{s,j,q}$ is a random variable which is generated by Rayleigh distribution, and $\alpha=3$ is the path-loss exponent. Channel power gains for fronthaul links are set to $h_{k_2}^{j,q}=\Omega_{k_2}^{j,q}{d_{j,q}}^{-\beta}$ where similar to access links, $d_{j,q}$ is the distance between RRH $j$ and BBU, $\Omega_{k_2}^{j,q}$ is a random variable generated according to the Rayleigh distribution, and $\beta=3$ is the path-loss exponent. The power spectral density (PSD) of the received Gaussian noise is set to $-174$~dBm/Hz. At each RRH, we set $P_{\text{RRH}_j}^{\text{DL}}=43$~dBm $\forall j\in\mathcal{J}$ and $P_{\text{RRH}_j}^{\text{UL}}=43$~dBm $\forall j\in\mathcal{J}$. For the BBU, we set $P_{\text{BBU}}^{\text{DL}}=47$~dBm, and for each user, we set $P_{\text{USER}_i}^{\text{UL}}=18$~dBm. The frequency bandwidth of wireless access and fronthaul links are $W_\text{AC}=5$~MHz and $W_\text{FH}=10$~MHz, respectively. Moreover, the bandwidth of each subcarrier is $W_S=625 \times 10^3$~Hz. The QoS exponent is $\theta=11$
\cite{ngo2014joint}. Unless otherwise stared, we consider 2 RRHs, 2 users in each RRH and 2 slices. In this section, we use Monte-Carlo method for simulation where the optimization problem is solved for 1000 channel realizations  and the total transmit power is the average value over all derived solutions.
\vspace{-0.5em}
\subsection{PD-NOMA versus OFDMA for the {TI} in Our Setup}
\begin{figure}[t]
	\centering
	\vspace{-1.2em}
	\includegraphics[width=0.38\textwidth]{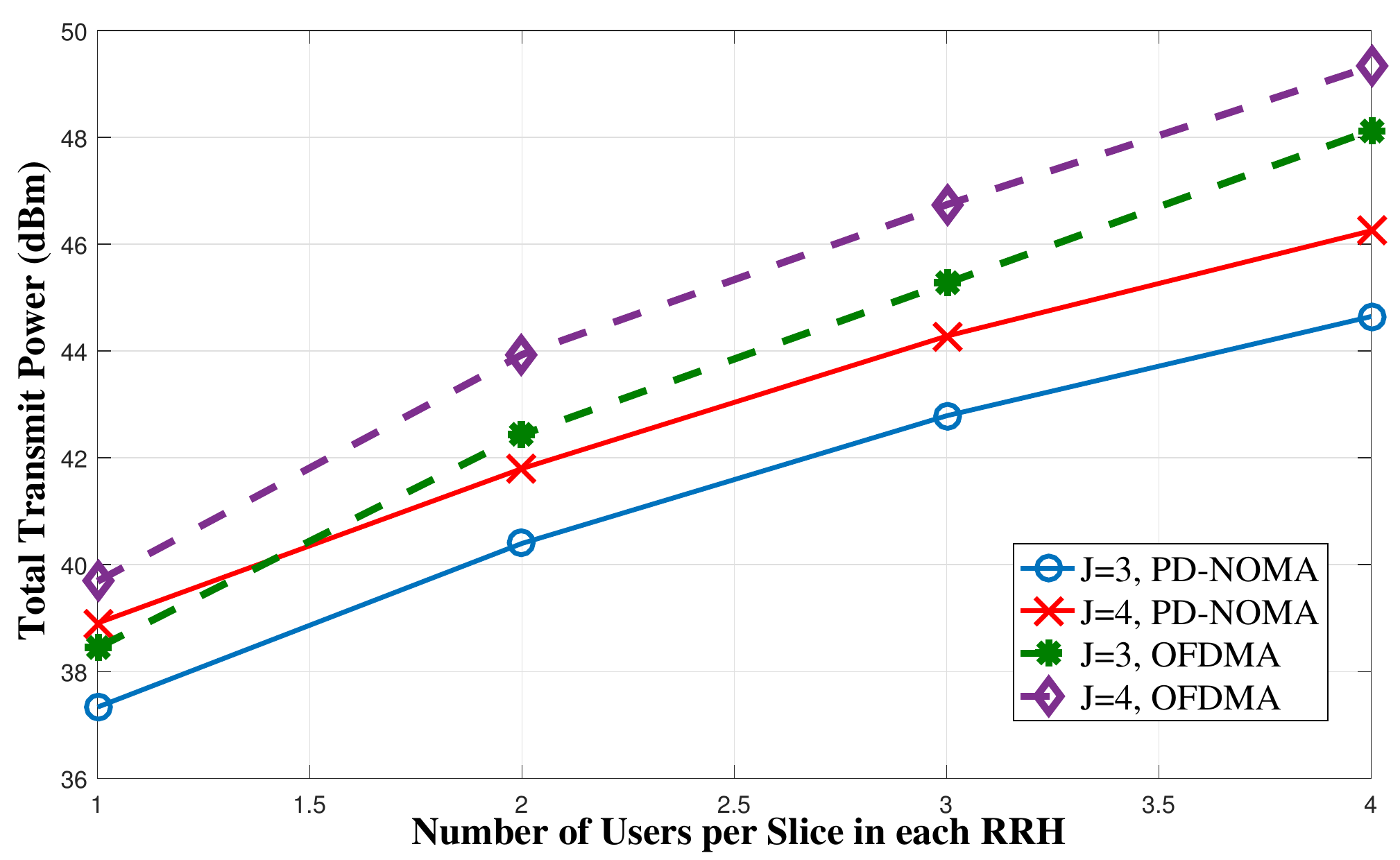}
	\caption{Comparing PD-NOMA and OFDMA versus number of users per slice in each RRH }
		\vspace{-1em}
	\label{Sim2}
\end{figure}

We start our study with comparing the performance of PD-NOMA and OFDMA for {TI} in the proposed problem. \textcolor{black}{For OFDMA-based system, problem \eqref{eqoa} by considering $L1 =L2 = 1$ is solved and simulated which consists of both dynamic power allocation problem, subcarrier allocation subproblem, and delay adjustment problem. Therefore, for OFDMA-based system the set of variables is similar to that in NOMA-based system. Indeed, the complexities of the two schemes are not the same to make the comparison exactly fair. Nevertheless, compared to OFDMA, NOMA only adds the SIC constraints "C4" and "C8" to the optimization problem, which are linear constraints and do not impose a major increase in complexity for solving the optimization problem.} Fig. \ref{Sim2} shows the total transmit power versus the number of users per slice in each RRH for these two cases. As expected, 
 the total transmit power increases by increasing the value of reservation rate  $R_{\text{rsv}}$.
\textcolor{black}{As shown in this figure, the NOMA-based system outperforms the OFDMA-based system. From Fig. \ref{Sim2}, increasing the number of users can significantly increase the total transmit power in the OFDMA-based system, e.g., the transmit power increases around 25\% per each slice in each RRH. This simulation reveals the efficiency of NOMA compared to OFDMA for the tactile users. Therefore, in the following figures, we only demonstrate NOMA-based simulation results.}

\vspace{-0.5em}
\subsection{Effects of Network Parameters}
Here, we study the effects of other network parameters on the performance of the proposed algorithm based on PD-NOMA assisted C-RAN.
\begin{figure}[t]
	\vspace{-0.5em}
	\centering
	\includegraphics[width=0.38\textwidth]{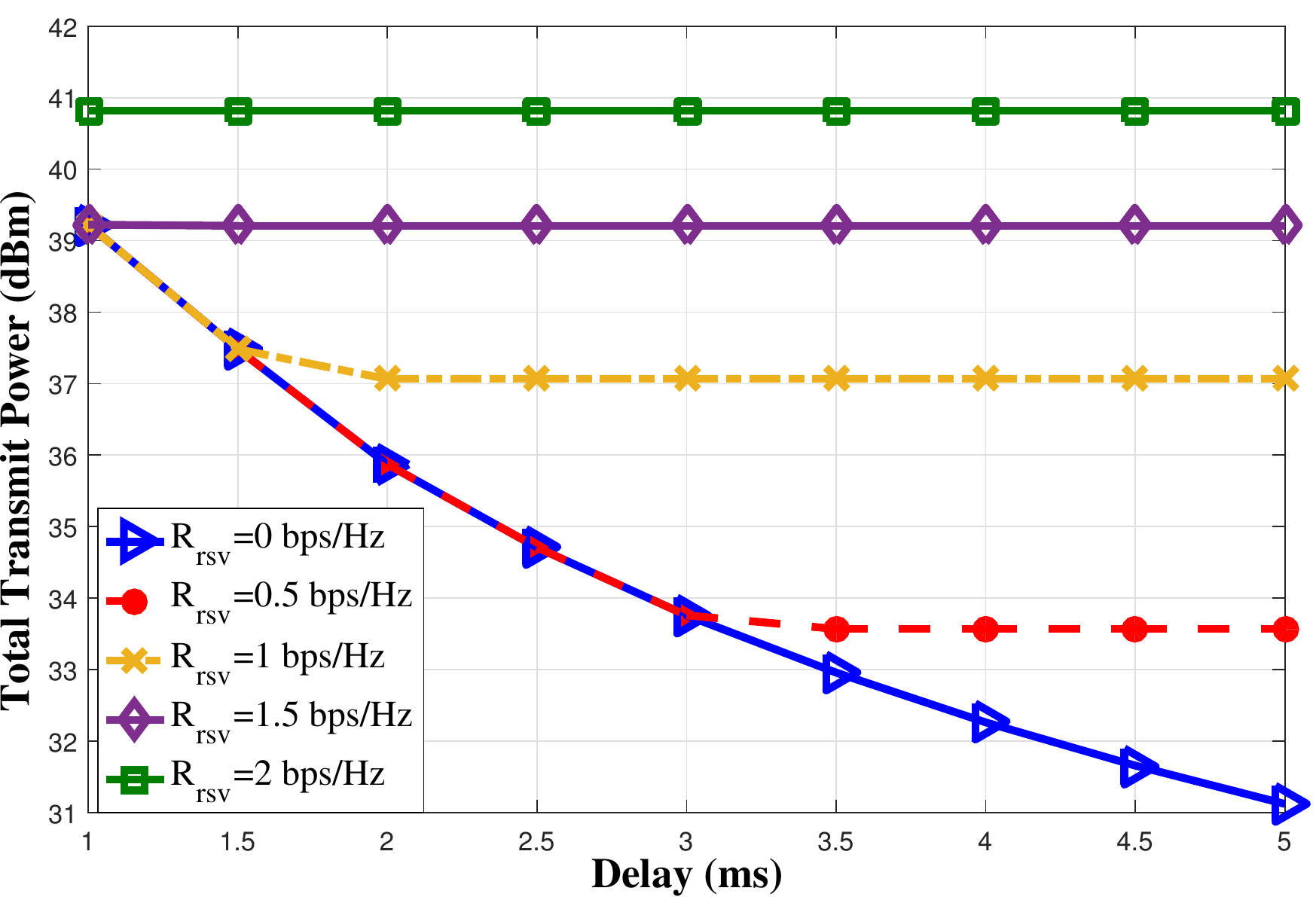}
		\vspace{-0.5em}
	\caption{Total transmit power versus delay for different $R_{\text{rsv}}$}
	\vspace{-1em}
	\label{Sim5}	
\end{figure}			   
\begin{figure}[t]
	\centering
	\includegraphics[width=0.38\textwidth]{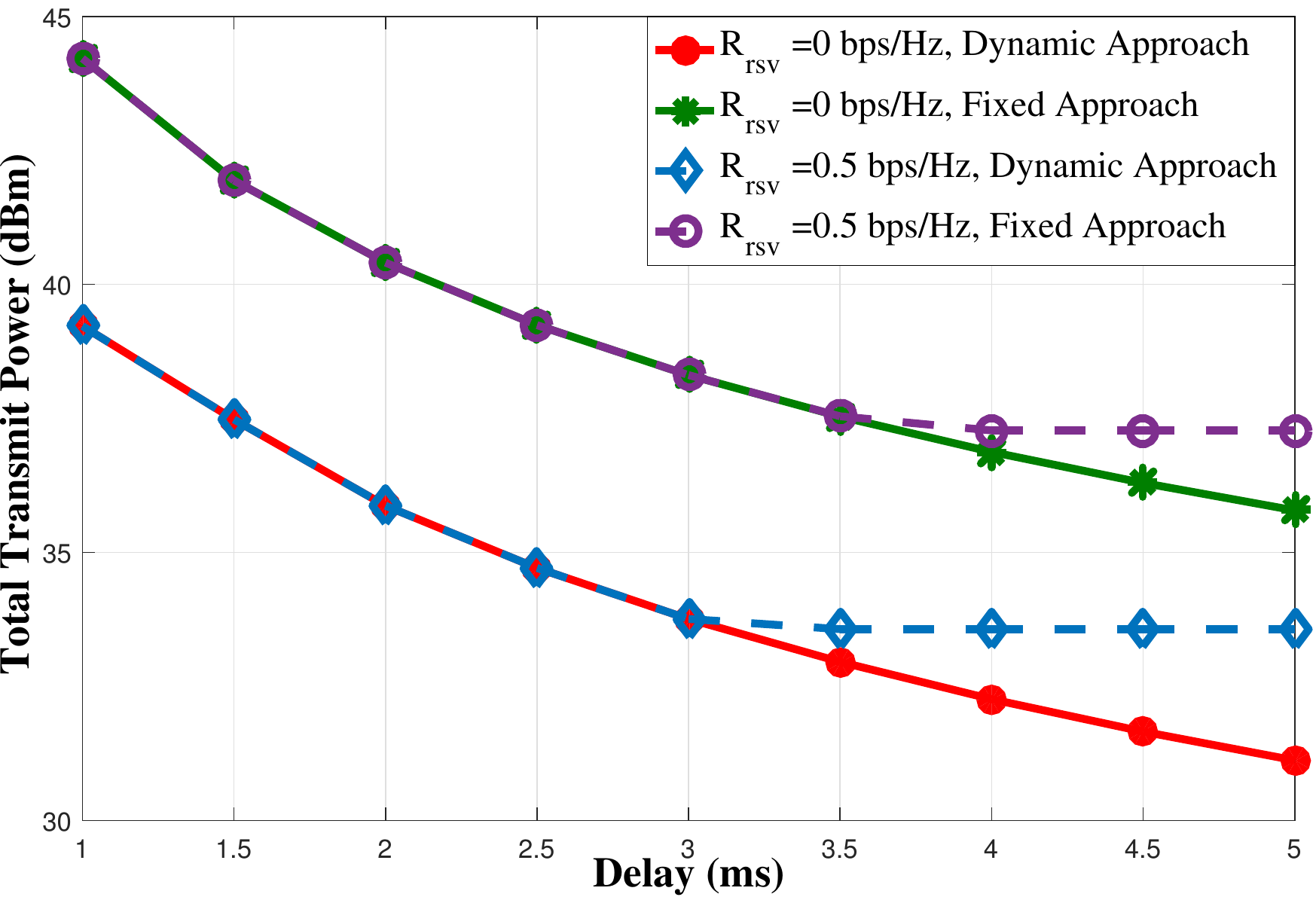}
		\vspace{-0.5em}
	\caption{Total transmit power versus delay for dynamic approach and fixed approach}	
		\vspace{-1em}
	\label{Sim6}
\end{figure}
To evaluate the performance of the proposed system model on a comparison basis, we consider a different scenario: in our proposed system model, we adjust the delay dynamically to minimize the transmit power which is called dynamic approach. In the new scenario  which is called fixed approach, we assume the delay constraints are fixed and cannot be adjusted for access and fronthaul links. In this case, we have a new optimization problem, referred to as, the relaxed problem,  in which we remove the delay variables from problem \eqref{eqoa} and ignore constraint C9. Moreover, we set access and fronthaul delay manually in constraints  C10, C11, and C12 as  ${D^j_{{\rm{max}}}}={D^{\text{BBU}}_{{\rm{max}}}}={D^{i,j}_{{\rm{max}}}}=D^{\text{Total}}_{{\rm{max}}}/3$. The new problem can be solved with the DC approximation.

\begin{figure}[t]
	\centering
	\vspace{-0.5em}
	\includegraphics[width=0.38\textwidth]{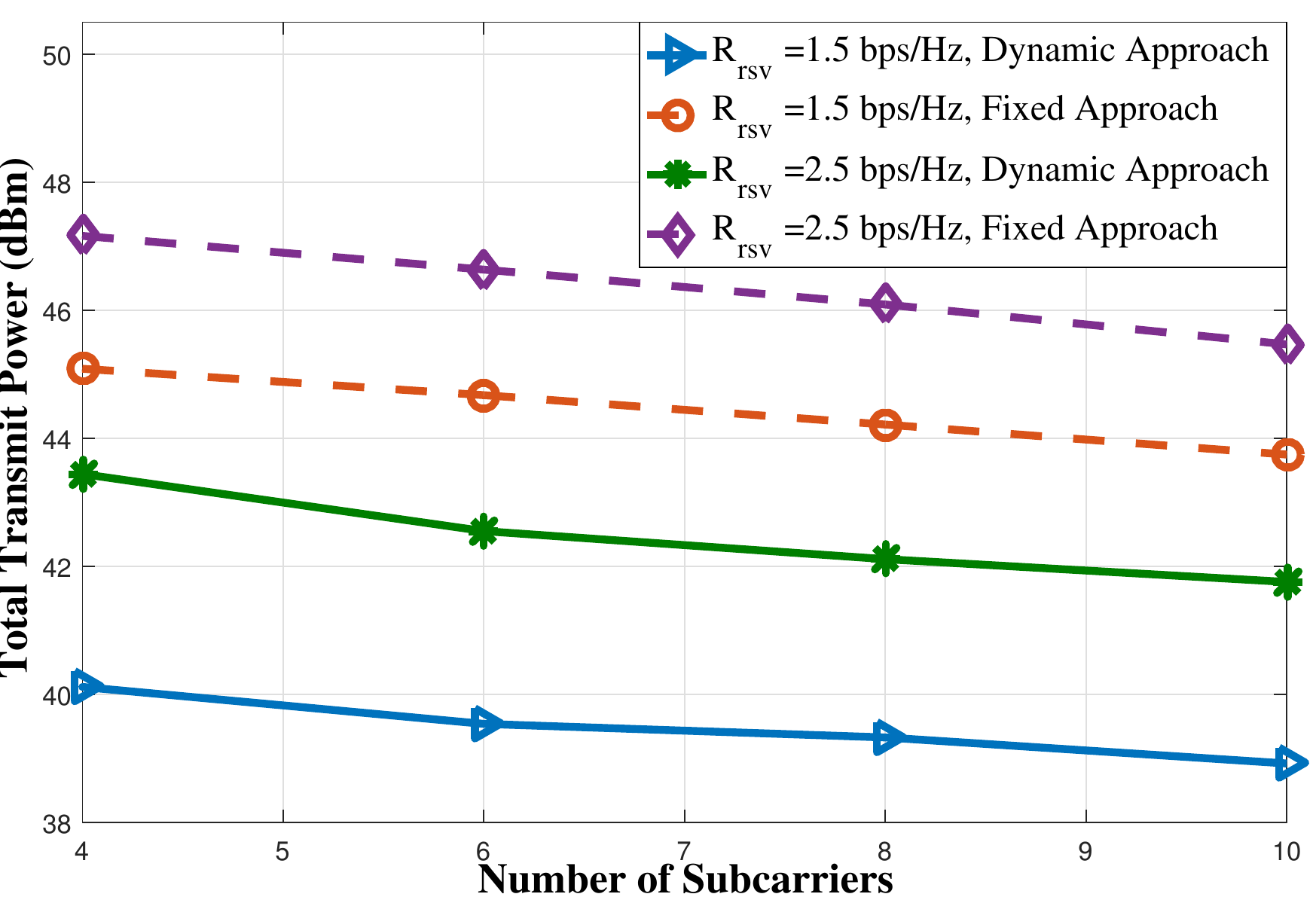}
	\vspace{-0.5em}
	\caption{Total transmit power versus number of subcarriers}
		\vspace{-1em}
	\label{Sim7}
\end{figure}

In Fig. \ref{Sim5}, we investigate the effect of the actual value of the delay $D^{\text{Total}}_{{\rm{max}}}$ on the total transmit power.
	As can be seen, for smaller values of reservation rate, every 1 ms decrease in delay requires 2 to 3 dB increase in the total transmit power. However, for larger values of reservation rate, the total transmit power is almost independent from the delay. In addition, it can be seen that by ignoring the reservation rate, the total transmit power decreases when delay increases.
 Moreover, from Fig. \ref{Sim6}, the proposed system model has a considerably better performance than the system model corresponding to the relaxed problem. As can be seen, by the dynamic adjustment of the delay, we can save around $5$~dBm in transmit power.
In Fig. \ref{Sim7},  the total transmit power versus the number of subcarrier for $D^{\text{Total}}_{{\rm{max}}}=1$ms. Fig. \ref{Sim7} is demonstrated where total transmit power decreases by increasing the number of subcarriers. 
Moreover, from Fig. \ref{Sim7}, up to 5 dBm transmit power can be saved.

\vspace{-0.5em}
\section{Conclusion}
In this paper, we proposed a novel queuing model for the {TI} services in  PD-NOMA-based C-RANs serving several pairs of tactile users. For each pair of tactile users within C-RAN coverage area, our setup includes RRH and BBU queuing delays in one E2E connection which is a more practical scenario in this context compared to previous works. We proposed a resource allocation problem to minimize the transmit power by considering E2E delay of joint access and fronthaul links for each pair of tactile users  where the delays of fronthaul and access links are dynamically adjusted to minimize the transmit power. To solve the highly non-convex proposed resource allocation problem, we applied SCA method. Simulation results revealed that by dynamic adjustment of the access and fronthaul delays, transmit power can be considerably saved compared to the case of fixed approach per each transmission.

\vspace{-0.5em}

\ifCLASSOPTIONcaptionsoff
\newpage
\fi

\bibliographystyle{ieeetran}
\bibliography{Tactile}		
\bibliographystyle{ieeetr}

\end{document}